\documentclass[journal]{IEEEtran}

\usepackage[utf8]{inputenc}
\usepackage[T1]{fontenc}
\usepackage{siunitx}
\usepackage[colorlinks=true, allcolors=blue]{hyperref}
\usepackage{amsmath}
\usepackage{amsfonts}
\usepackage{soul} 

\usepackage{tikz,xcolor}
\definecolor{lime}{HTML}{A6CE39}
\DeclareRobustCommand{\orcidicon}
{
    \begin{tikzpicture}
    \draw[lime, fill=lime] (0,0) circle [radius=0.16] 
    node[white] {{\fontfamily{qag}\selectfont \tiny ID}};    \draw[white, fill=white] (-0.0625,0.095) circle [radius=0.007];    
    \end{tikzpicture}
    \hspace{0mm}}
\foreach \x in {A, ..., Z}{%
    \expandafter\xdef\csname orcid\x\endcsname{\noexpand\href{https://orcid.org/\csname orcidauthor\x\endcsname}{\noexpand\orcidicon}}
}

\usepackage{url}
\usepackage{listings}
\usepackage{color}
\usepackage{xcolor}
\usepackage{caption}
\usepackage{pgfplots}
\definecolor{dkgreen}{rgb}{0,0.6,0}
\definecolor{gray}{rgb}{0.5,0.5,0.5}
\definecolor{mauve}{rgb}{0.58,0,0.82}
\usepackage{makecell}

\begin{document}

\title{Design of VR Engine Assembly Teaching System}

\author{
    \IEEEauthorblockN{Jiayu Zhang}
    \IEEEauthorblockA{\\Department of Computer Science and Technology, Southern University of Science and Technology
    \\Shenzhen, China
    \\11812425@mail.sustech.edu.cn}
}

\markboth{Journal of \LaTeX\ Class Files,~Vol.
}%
{Shell \MakeLowercase{\textit{et al.}}: Bare Demo of IEEEtran.cls for IEEE Journals}

\maketitle

\begin{abstract}
  Virtual reality(VR) is a hot research topic, and it has been effectively applied in military, education and other fields. The application prospect of virtual reality in education is very broad. It can effectively reduce labor cost, resource consumption, stimulate students' interest in learning, and improve students' knowledge level. New energy vehicles have also been widely promoted in recent years, and the production of new energy vehicles has played a key role in it. However, the teaching of car engine disassembly and assembly still retains a more traditional way. That's why applying VR technology has high significance. This project uses the Unity 3D engine to develop a VR-based engine teaching software, which aims to allow users to use VR headsets, handles and other accessories to simulate the disassembly and assembly of car engines in a virtual environment. We design a modular system framework and divided the software into two layers, the system layer and the function layer. The system layer includes a message system and a data configuration system. The functional layer includes the user interface system, disassembly and assembly function, and data module. In addition to fulfilling functional requirements , we used the Unity UPR tool to check out performance issues, and optimized product performance by turning off vertical sync and turning on static switches for some scene objects.
\end{abstract}


%
\IEEEpeerreviewmaketitle

\section{Introduction}

\subsection{Background}

With the rapid development of information technology, the research of virtual reality has gradually become a hot topic, and it has also been effectively applied in military, education and other fields. In the field of education, virtual simulation software is also widely used, and the teaching significance is very far-reaching. On the one hand, the simulation experiment teaching through software can effectively reduce labor costs and reduce the consumption of equipment resources. On the other hand, it can also help improve students' sense of participation and immersion, fully stimulate students' interest, and effectively improve students' knowledge level.

The teaching of automobile engine disassembly and assembly still retains the traditional theoretical teaching mode, and the supporting practical courses are relatively scarce. Therefore, students usually cannot fully apply theoretical knowledge to practice, nor can they achieve the goal of consolidating and improving theoretical level in practice.

In order to make up for the shortage of automobile engine disassembly and assembly teaching in reality, and give full play to the application significance of virtual reality technology in the field of education, this project uses the Unity 3D engine to develop engine disassembly and assembly teaching software based on virtual reality. In addition to meeting the basic functional requirements, the performance data of this project has also been optimized to meet certain standards to ensure that users have a good experience.

\subsection{Related Work}

\subsubsection{Current situation of automotive industry teaching}

With the development of science and technology, the automotive industry is also changing rapidly, and various models of engines are emerging. Vocational colleges in general need to focus on practice, and in order to keep up with the changes in the automotive industry \cite{lan2017development,xu2019online}, they need to regularly update the equipment needed for practical teaching, which requires adequate financial support, and many vocational colleges are currently facing the problem of insufficient funding\cite{AiLian2011Vocational}.

The automotive industry usually requires the use of many bulky professional equipment, such as frame beam aligners, complete vehicle painting rooms, etc. This leads to a particularly high demand for teaching space. Classroom resources in general vocational colleges are also very limited, and often students are not supported by ideal teaching space \cite{2017Application}. In addition, even if there are enough teaching resources, there are still many security risks in the process of teaching practice \cite{Tingting2021VR}.

Due to all these difficulties, teaching in vocational colleges in the automotive industry is usually based on purely theoretical teaching, and practical teaching is very scarce. This leads to boring teaching contents, lack of interaction between students and teachers, and serious lack of interest and concentration of students in class. Students are often completely absent from the classroom, making it very difficult for teachers to manage \cite{Juan2014VR}.

\subsubsection{Research status of virtual simulation teaching software}

When virtual reality technology was first introduced, users needed to use many hardware devices to achieve an immersive experience, but factors such as blurred virtual images, low real-time performance, and a wide variety of hardware led to its not being universally applied and developed \cite{lan2018directed,DeFu2017Application}. Nowadays, with the rapid development of information technology \cite{lan2022semantic,gao2021neat}, the limitation of hardware devices for virtual reality has been broken, and the display effect of computers has been significantly improved, so virtual reality has entered the public's view again \cite{MinJie2011Research}.

At present, the teaching software of virtual reality simulation has been widely used in western countries \cite{Ferrero1999simulation}. 
The virtual world working group jointly established by New Zealand and Australia and Second Life (SL) researched by Linden Laboratory in the United States are typical representatives. 
In the United Kingdom and the United States, SL projects have a very high market share \cite{su2014case}. 
Since the launch of SL, the number of internal users has grown to 37 million \cite{Potkonjak2016Virtual}, with over 300 universities around the world, using SL on teaching courses or studying user behavior on SL \cite{Siemens2004Connectivism}. 
Some special industries have strong demand for virtual simulation teaching software, such as pilot driving training, historical scene review \cite{Xiao2018Scene} and so on  \cite{lan2019simulated,lan2019evolutionary}.

For the automotive industry, there are also some practical developments and applications \cite{lan2018real,lan2019evolving}. For example, North China University of Science and Technology Fan Luning developed a VR technology-based auto repair teaching and training system, Paris National Mining High School developed a VR training system for disassembly and assembly of complex mechanical equipment  \cite{Chao2017survey}.

\subsubsection{Comparison between traditional teaching and VR teaching}

This project is to develop a VR teaching software to replace the traditional engine disassembly and assembly teaching mode to achieve better teaching effect, so we compared between traditional teaching and VR teaching in various aspects in \autoref{comparison}, such as the teaching venue, and the difficulty of detecting the knowledge mastery after learning. From this comparison we can get the conclusion that VR teaching has such many advantages over traditional teaching that it is important to apply VR to vocational teaching.

\begin{table}[!ht] \small \centering
        \caption{Comparison between traditional teaching and VR teaching}\label{comparison}
        \begin{tabular}{p{1.8cm}|p{2.5cm}p{3cm}} 
        \hline 
            ~ & Traditional teaching & VR teaching\\ \hline 
            Footprint & require more equipment, occupy a larger area \cite{Feng0AR} & only require VR equipment and computer, smaller area \\ \hline 
            Teacher-student ratio & class of 15-30 people, two teaching teachers \cite{KeRong2020Thoughts} & everyone can follow the instructions in software to learn \\ \hline 
            Safety issues & practical operation may occur safety accidents \cite{Tingting2021VR} & relatively safe, almost no safety problems\\ \hline 
            Inspection difficulty & requires teacher supervision, inspection is more difficult \cite{MinJie2011Research} & software can detect correctness of students' operation in real time\\ \hline 
            Students' interest in learning & traditional teaching methods are low in learning & novel and unique teaching methods, students are in high interest in learning\\ \hline 
        \end{tabular} 
\end{table}

\subsection{Research content and structure}

The main research object of this paper is virtual reality technology, and the application of virtual reality technology in the vocational skill training of engine disassembly and assembly is discussed. Firstly, the requirements of the teaching software for disassembly and assembly of the virtual reality engine are discussed and analyzed. Then the system architecture of the software is designed based on the analysis of requirements, and the functional requirements are developed in a modularized manner. Finally, the software is tested for function and performance. Based on the test results, the software is modified and the performance differences before and after optimization are compared, so that the performance reaches the standard. The structure of this paper is as follows:

The first chapter is an introduction, which mainly introduces the project background and significance, the teaching status of engine disassembly and assembly, the research status of virtual simulation teaching software, and the innovation points of the paper;

The second chapter is the design and development of the virtual reality engine disassembly and assembly teaching software system, which fully expounds the design and development process of the software system, including requirements analysis, system architecture design, and specific implementation of key functions, etc.;

The third chapter is the test and optimization of the virtual reality engine disassembly teaching software system;

The fourth chapter is the conclusion, which includes a summary of the work done during our project and an outlook on the application of virtual reality technology in the field of education.

\subsection{Innovation}

The simulation teaching software based on virtual reality has been widely studied. Based on reading literature and researching open source software, this paper summarizes with its own ideas, and designs a virtual reality engine disassembly teaching software system based on the requirements. The innovations of this paper are described as follows:

\begin{enumerate}
    \item This paper proposes the design and development of teaching software for engine disassembly and assembly based on virtual reality for vocational college education in the automobile industry, which has a certain educational significance for vocational education in the automobile industry, and improves the learning effect of students while reducing education costs;
    \item This paper expounds the design of the system architecture of the engine disassembly and assembly teaching software based on virtual reality, and develops the functions of the user interface, assembly system, task system and other functions in a modular way in the designed system;
    \item This paper conducts a complete functional test and performance test of the engine disassembly and assembly teaching software based on virtual reality, and makes a clear direction adjustment for the unsatisfactory performance data, and finally makes the performance data meet the standard.
\end{enumerate}

\section{System Design and Implementation}

\subsection{Software requirements analysis}

When designing and developing a software, first of all, it is necessary to clarify the requirements of the software. The ultimate goal of this project is to design a teaching software for the disassembly and assembly of Buick Willam automobile engines based on virtual reality. There are two types of groups - one is to use this software for learning, including vocational schools students in the automotive industry and people who study independently for interest, industry needs or other reasons. Their requirements are generally based on the purpose of learning. On top of the basic requirements of learning basic knowledge, sometimes they also require that the process of learning be interesting enough. The other type is teachers who use this software for auxiliary teaching. Their requirements are generally based on the purpose of convenient use and teaching students. In the case of ensuring the correct teaching knowledge, it is also convenient to adjust the teaching knowledge in the software. Various parameters should be able to be adjusted, such as a component ID, the task sequence and so on \cite{lan2022time,lan2021learning2,lan2021learning1}. Below we provide a more detailed analysis of the needs involved in these two groups.

For the groups who use this software for learning, we analyze their requirements as follows. First of all, we need to ensure that users learn the basic knowledge correctly. For this reason, we need to design a reasonable system architecture to ensure the modular separation of functions, to ensure that there are no errors during the running of the program, and to run correctly according to the configured parameters. Secondly , the user require to keep the software running smoothly during the use process, which requires the use of some technologies in the development process to ensure a stable running frame rate and reduce memory consumption, so that the program can also have high quality on devices with lower configuration. Finally, during the user's learning process, the art quality of the software will also affect the user's learning effect. The art quality generally includes model accuracy, user interface design, texture restoration, etc. These should also be guaranteed during the development process.

For the groups who use this software for teaching, we analyze their requirements as follows. The general purpose of such groups is to use software to assist their original teaching and improve students' learning enthusiasm and learning effect. In the teaching process, it is inevitable that there will be some data adjustments, such as replacing the parts used, changing the order of task composition, etc. Here, developers need to leave an entry for easy data adjustment during the development process, rather than writing fixed code. , so that teachers can easily adjust in the editor.

\subsection{System architecture design}

In the requirements analysis, we mentioned that the most basic requirements of users are satisfied that the software has a good system architecture, which has a modular function design, which can make the function boundaries clear in the development version, and locate the problem in time when there is a problem. In our teaching software, the system architecture is mainly divided into two layers, the system layer and the function layer. The system layer contains some low-level codes that are not related to the project, including the message system and data configuration system; the functional layer includes some business codes that are strongly related to the business requirements of the project, including user interface, disassembly logic, process control, etc. .

\begin{figure}
    \centering
    \includegraphics[width=.4\textwidth]{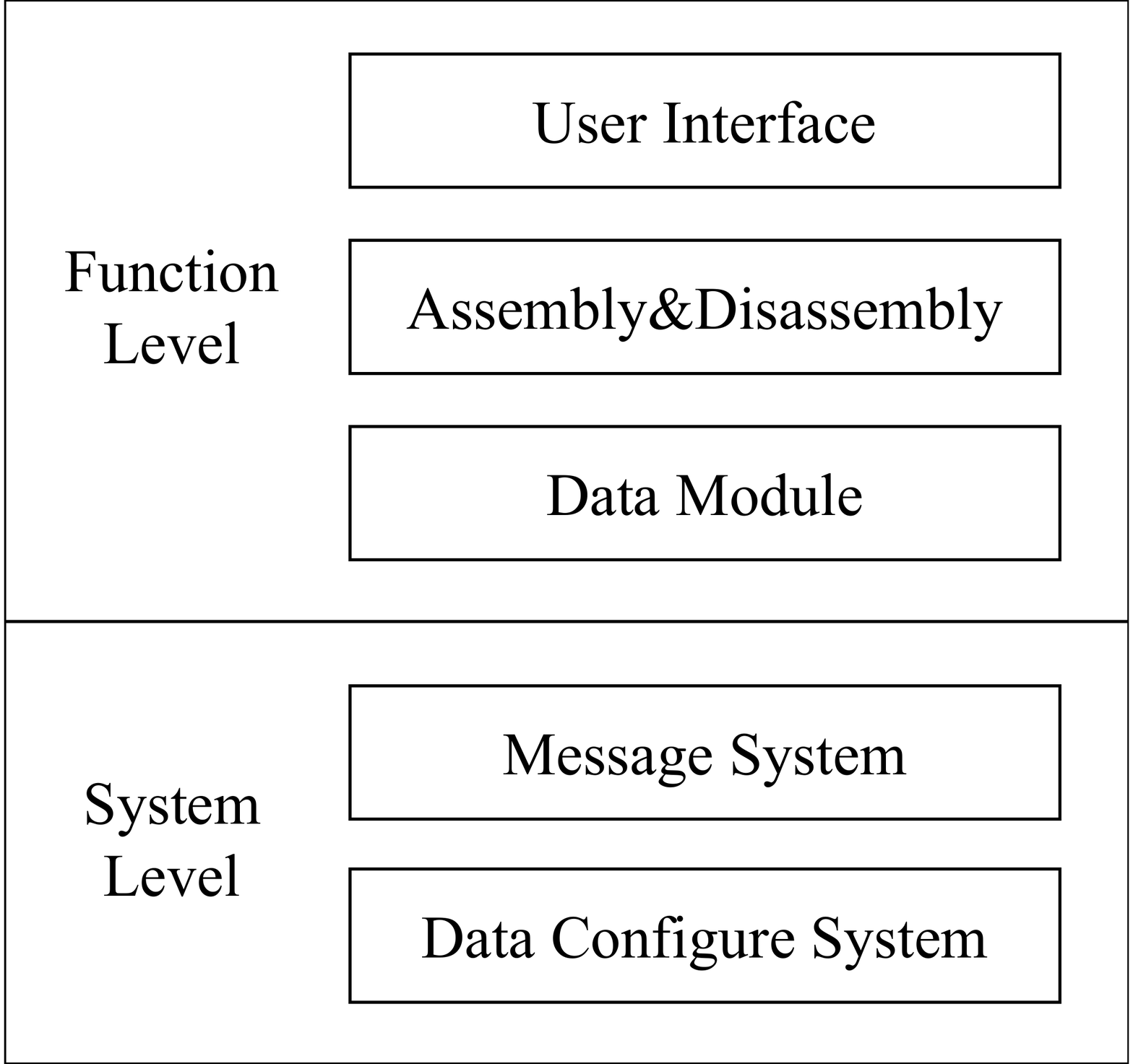}
    \caption{System Architecture}\label{system-design}
\end{figure}

\subsubsection{System layer design}

We start from the system layer, build the underlying framework for the project, and lay a solid foundation for the business development of the project.

First of all, we need to implement the message system to realize the interaction between objects and avoid code coupling. We design a custom type of message through inheritance registration, and carry the necessary parameters in the message to pass to other objects. When an object that needs to be notified of a message is generated, register and monitor the necessary message type. If a message is sent by another object in the monitored message, the registered callback will be called, and the specific command will be executed using the parameters carried in the message.

Secondly, we need to standardize the configuration method of data and provide the basic data development entry, such as the combination of tools and the process of tasks need to use configuration files. CSV files are most suitable for such configuration files . CSVs are plain text files that make data exchange easier and format conversion easier. We need to develop a general CSV file conversion system that reads CSV files into custom classes and wraps the data into corresponding objects.

\subsubsection{Function layer design}

At the functional layer, we develop the main business requirements to implement the functionality of the project itself.

The first is the user interface system (UI system), which consists of two major parts. One is the pad system, which contains the main non-intermediate interaction logic, including mode selection, task selection, etc. The tablet acts as a Operable objects can be picked up and put down at will in the virtual reality space. The other is a large-screen system. The main function of the large-screen system is that users can observe their own task progress in real time when they are completing tasks. User can also see real-time score in training mode.

The second is the assembly function, which consists of two parts. One is the component system, which controls some unique behaviors of the parts, such as colliders, rigid bodies, grabable scripts, highlight effects, animation controls, etc.. The other is the tool system. This part uses the configuration file to define the combination logic between tools and tools, the interaction logic between tools and parts, and also specifies that the parts need to configure the tool conditions for disassembly and assembly of the used tools.

Finally, there is the data module, which specifies the content and process control of the task, and also includes a scoring system, which controls the closed-loop operation of the entire project at the top of the functional layer.

Next, this paper will select several key modules and describe their design and implementation.

\subsection{Implementation of key modules}

\subsubsection{Assembly function realization}

The assembly function is the core function of the project, which includes component animation, component state control, tool combination, interaction conditions and interaction effects between components and tools, and so on. For components and tools, there are some general logics, such as grabbing and placing, highlighting, prompts, collision interaction, etc.. But they also have some unique logics. Components need to configure the assembly conditions, and tools have a combination logic. The following is the script logic common to components and tools:

\begin{enumerate}
    
    \item \textbf{Collider:} Collider, a switch used to detect collisions between objects and objects, objects and hands, and realize parts disassembly, tool combination, etc.;
    \item \textbf{Advanced Interactable Obj:} An interactive object component inherited from VRTK. It is the core script for object control, controlling the collision, grabbing, using of objects. On the basis of the parent class, a one-click interactive switch is added and ray interaction is allowed;
    \item \textbf{Panel Spawn Tooltip Trigger:} Control the display and hide of the object prompt page to help users disassemble and assemble.
\end{enumerate}

The following will explain the realization of the assembly function from the unique functions of the part system and the tool system.

(1) Component system implementation

For a fixed model engine, its composition generally does not change. So we directly edit its script and configure properties for each component model. For components, they can be divided into two categories: tool-dependent and non-tool-dependent. Tool-dependent components refer to components that need to be operated with tools of a fixed type, while non-tool-dependent components are components that can be operated without tools. Some of these components have preconditions, and operations cannot be performed if the preconditions are not completed. The part-specific scripts are as follows:

\begin{enumerate}
    
    \item \textbf{DA Obj Ctr:} Components control script, mainly divided into two sub-categories according to whether the tool is dependent or not: VRDA Screw and VRDA Parts Ctr, which mainly control the operability of parts, the playback of disassembly animation, state control, etc.;
    \item \textbf{Disappear Effect:} The disappearance effect of the object, which controls the movement direction, distance, duration, etc. of the object when the parts are disassembled.
\end{enumerate}

If the tool-dependent components depend on the combination of wrench tools, an additional Wrench DA Script script needs to be added, as shown in \autoref{wrench-script}, in the property interface of the editor to control the type of tools used and other details required for disassembly.

\begin{lstlisting}[language=Java,caption={Wrench Script Class},label=wrench-script,,captionpos=b]
public class WrenchDAScriptV2
{
    public Transform wrenchUsePos;
    public bool matchPosDirection = true;
    public WrenchUseCondition wrenchUseCondition;
    public ScrewOutLevel screwOut = ScrewOutLevel.TwoCM;
    public float customOutLevel;
    public bool autoFix;
}
public struct WrenchUseCondition
{
    public string banshouID;
    public string fixBanshouID;
    public string jieganID;
    public string taotongID;
    public bool needJiegan;
    public int minFixTorsionRange;
    public int maxFixTorsionRange;
}
\end{lstlisting}

(2) Tool system implementation

The tool system differs from the part system in that it uses a relatively high repetition rate and relatively low coding complexity. But because of this, it needs to be more versatile. In the project, most of the tools used are combinations of wrenches, posts and sockets. Other tools have special script control. We only focus on the former, which is the commonly used combination tools. For wrench, there are the following script components:

\begin{enumerate}
    
    \item \textbf{VR Wrench Ctr:} mainly controls the wrench itself, such as the combination of data between tools, the update of display data, the playback of animation, etc.;
    \item \textbf{Wrench Combine Ctr:} controls the combination logic of wrench, post and socket.
\end{enumerate}

For a post or a wrench, it is represented as a combinable tool by adding the Combine Able Wrench Parts component, and the combination logic of the combined tool is controlled by this component. All the combinable tools are placed in the toolbox, and other special tools are placed together with the toolbox on the tool table for users to access at any time when the program is running.

The tools vary from model to model, and so do the combinations. For ease of development and configuration, we use CSV files to define composition rules between control tools. For a wrench, its Kit property defines a list of combinable tool ids. During the actual combination process, only the tools in this list can be combined with the wrench.

\subsubsection{Task module}

\begin{figure*}
    \centering
    \includegraphics[width=.9\textwidth,trim=50 400 50 0,clip]{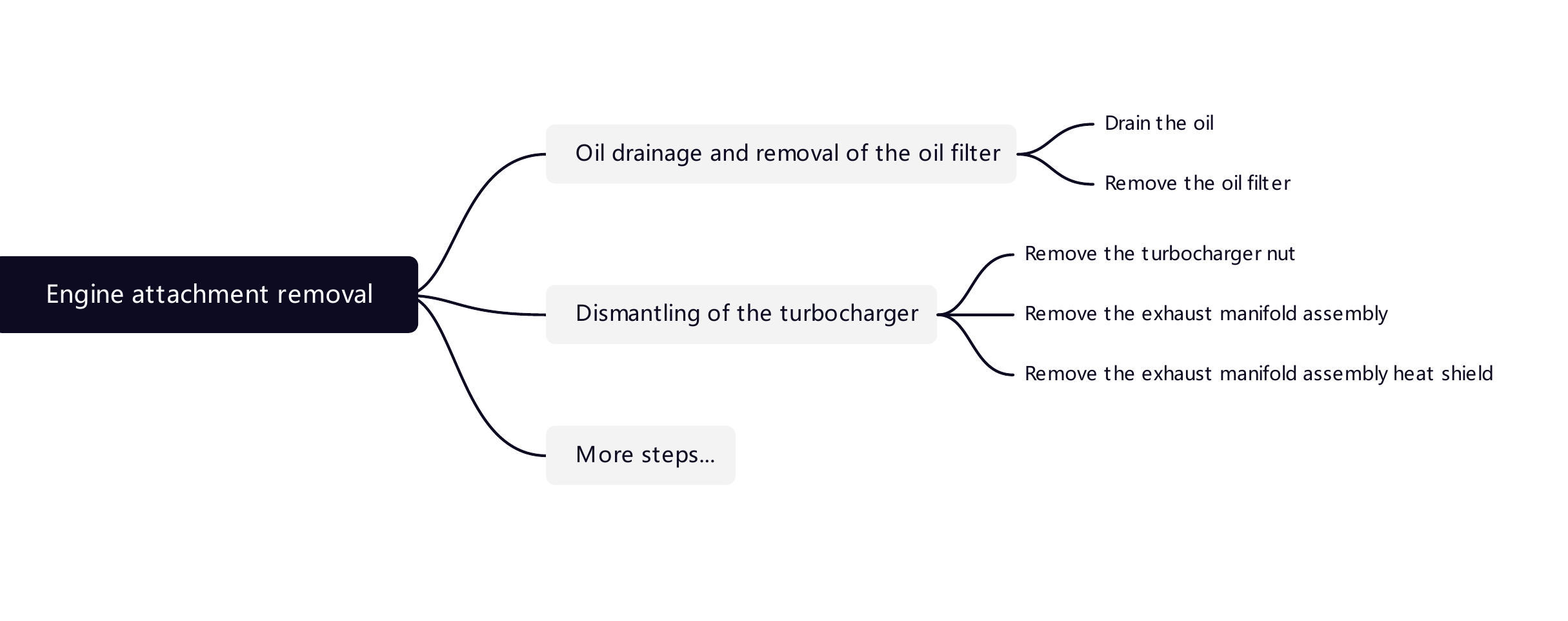}
    \caption{Task Composition}\label{task}
\end{figure*}

This project is divided into three modes in total. For each mode, the operations that the user needs to perform are consistent, that is, to perform step-by-step operations with a single part as the granularity under the guidance of the task, the difference is only with or without guidance and real-time scoring . Starting from the requirements, the entire assembly process is divided into several large tasks, each task is divided into step groups, and the step group is composed of some small steps.

The first big task of disassembly is shown in \autoref{task} - engine attachment removal. The task of engine attachment removal can be broken down into eleven step groups, the first two of which are shown in the figure. Further subdivisions are steps. The first few steps we complete the task of disassembling the engine attachments are draining the oil, removing the oil filter, removing the turbocharger nut, removing the exhaust manifold assembly and removing the exhaust manifold assembly heat shield. The contents of these tasks are configured in the csv file.

In the task module, we record data such as the completion of steps in the running process, error operations, scores, etc., and also provide many functions such as initializing task data, checking the completion of steps, and submitting tasks for use by other modules. 

When the task is selected through the pad, the task module obtains the task id that needs to be initialized from the global, all objects determine their own state according to the task, and provide the user with the starting scene of the task. During the running process, whenever the state of the part changes, a state change message will be sent, and the progress information of the task module will be updated in the registered callback. When submitting a task, calculate the user's final score and restore the task-related information of the entire scene.

\section{System Testing and Optimization}

\subsection{Testing target}

\begin{enumerate}
    
    \item Verify the functional integrity and correctness of the virtual reality engine disassembly and assembly teaching software;
    \item Verify the performance requirements of the virtual reality engine disassembly teaching software.
\end{enumerate}

\subsection{Testing method and content}

\subsubsection{Function testing}\label{sec:function-testing}

The functional test is mainly used to verify the functional integrity and correctness of the virtual reality engine disassembly and assembly teaching. The test method is to package and release the software, use VR equipment to simulate real usage scenarios, test the functions defined by the requirements one by one, and test the functions. We need to test each function about whether it is included and the running result is as expected. The key required function points are as follows:

\begin{enumerate}
    
    \item Pad can normally jump to the corresponding page and execute the corresponding logic;
    \item The large screen can normally display the completion progress of the steps;
    \item The tool can be assembled normally;
    \item Components can be disassembled and installed normally;
    \item Tools and components display correct specular effects;
    \item The task steps can be carried out normally;
    \item Auxiliary functions such as video learning and general engine animation can run normally.
\end{enumerate}

\subsubsection{Performance testing}

According to the virtual reality experience standard technical white paper, the user's immersive experience quality, video viewing experience, and interactive experience quality are all important components of the quality of virtual reality products \cite{lan2016developmentVR,lan2016developmentUAV}. Immersive experience quality, including screen resolution, running frame rate, etc., is an important performance criterion for us.

Most of the standards such as screen resolution are determined by the hardware, and the running frame rate is an indicator that we can mainly test and optimize if there are problems. Generally speaking, VR products can only give users a better experience when they reach 90fps. In addition, if there is a performance indicator that is obviously not achieved, even if the software is not significantly stuck in subjective experience, this indicator needs to be optimized.

We use Unity UPR, a professional performance detection tool officially launched by Unity, to test software performance. When we run the packaged software and open the UPR desktop version, we can upload the running information to the server during local testing, and then view the performance information on the web page.

\subsection{Hardware environment}

The environment for testing performance is as follows:

\begin{enumerate}
    
    \item \textbf{Desktop:} intel 8-core i7-8700 CPU, 16G memory, NVIDIA GeForce GTX 1080 discrete graphics;
    \item \textbf{VR equipment:} HTC Vive head-mounted display device and accessories, HTC Vive infrared base station.
\end{enumerate}

\subsection{Testing data}

\begin{table}[!ht] \centering \small
    \begin{center}
        \caption{UPR Test Data}\label{test-data}
        \begin{tabular}{l|c}
        \hline
            Performance point & Value \\ \hline
            Average Frame Rate & 68.52 \\
            Mean Frame Time & 15.06ms\\
            Memory Reserved Total Peak(MB) & 1523.583 \\
            DrawCall Peak(times)      & 880\\
         \hline
        \end{tabular}
    \end{center}
\end{table}

We test the integrity and correctness of system functions according to the test cases in section \ref{sec:function-testing}. For each use case, we run multiple rounds of testing. The test results show that the functions involved in each use case have been fully implemented, and the results produced by executing the use case are as expected.

We use Unity UPR to test the packaged program, and the key performance data obtained are shown in \autoref{test-data}. We found that the performance of the project is relatively good, the memory consumption is kept at about 1.5GB. The project running process is relatively smooth, and there is no lag. But there are also problems. The average running frame rate remains at around 68FPS, which has not met expectations. We pick two entry points:

\begin{enumerate}
    
    \item The Vsync module consumes a relatively long time, and this module is related to the vertical synchronization switch of the program;
    \item There are too many DrawCalls. This function is related to the rendering logic inside Unity.
\end{enumerate}

In the following, we will optimize from these two perspectives and analyze the performance data before and after optimization.

\subsection{V-Sync optimization}

Vertical Sync (V-Sync), from the perspective of the display principle of CRT monitors, the entire screen is composed of pixels. For each display screen, horizontal pixels are generated from left to right, and then vertical pixels are generated from top to bottom to form a complete screen. The graphics card digital-to-analog conversion circuit (DAC) controls the refresh rate of the display, and the graphics card DAC generates a vertical synchronization signal after scanning a frame. Turning on VSync in the game or engine editor means passing the VSync signal to the graphics card for 3D graphics processing, so that the VSync signal restricts the generation of the graphics card's 3D graphics to avoid screen tearing.

There are three options for vertical sync in the Unity editor:

\begin{enumerate}
    
    \item \textbf{Don't Sync:} Do not turn on vertical synchronization, and directly let the display start drawing after the graphics card draws a picture;
    \item \textbf{Every V Blank:} Turn on vertical sync and synchronize on every vertical sync signal;
    \item \textbf{Every Second V Blank:} Turn on V-Sync and synchronize with the V-Sync signal every second.
\end{enumerate}

By default, the Unity editor selects Every V Blank, making the program's frame rate completely controlled by the vertical sync signal. When this project is running, the content in the field of view does not change rapidly. Turning off the vertical sync will not cause screen tearing. Turning on the vertical sync will cause the program to wait for the vertical sync signal to occupy excess CPU during the running process. Therefore, choose to turn off the vertical sync. At the same time, it is necessary to specify a frame rate higher than expected in the code, so that the program is really not controlled by the vertical sync signal when it runs.

We test the applications before and after the vertical synchronization is turned off. VS\_On means to turn on the vertical synchronization, and VS\_Off means to turn off the vertical synchronization. The test data is shown in \autoref{vs-test}.

\begin{table}[!ht] \centering \small
    \begin{center}
        \caption{V-Sync Optimization}\label{vs-test}
        \begin{tabular}{l|ccc} 
        \hline
            Case & \makecell[c]{Maximum frame\\time/ms} & \makecell[c]{Average frame\\time/ms} & \makecell[c]{Average frame\\rate} \\
        \hline
            VS\_On       & 470.21    & 15.06 & 65.61\\
            VS\_Off      & 445.72    & 3.94  & 264.28\\
        \hline
        \end{tabular}
    \end{center}
\end{table}

We found that after turning off VSync, both the highest frame time and the average frame time were reduced, especially the average frame time was reduced by 74\%, which greatly optimized the performance. At the same time, the average number of frames has increased by 300\%, and the smoothness of the picture has also been greatly improved. It can be seen that turning off vertical synchronization is ideal for the performance optimization of this project.

\subsection{DrawCall optimization}

In Unity, the CPU prepares data such as models, materials, and textures, and the process of passing these data to the GPU to draw the picture becomes a DrawCall. In order for the CPU and GPU to work in parallel, there is a command buffer, the CPU adds commands to it, and the GPU reads commands from it. The CPU addition command needs to prepare the rendering object, set the rendering state, output the rendering primitive and other calculations. If too many DrawCalls are used to draw the same page, it will bring huge computing pressure to the CPU and affect the performance of the program. We found from the performance data that the number of DrawCalls in the program is too many, so we need to reduce the number of DrawCalls.

Using batching techniques can reduce the number of DrawCalls very well. Batch processing refers to processing multiple objects at a time, and only objects using the same material can be processed in batches. Batch processing in Unity is divided into dynamic batch processing and static batch processing: dynamic batch processing is basically automatically processed by the engine. As long as it is a dynamic object that can be moved and transformed, the engine will determine whether to perform dynamic batch processing, but it can really be batched. The processed objects need to meet many requirements such as the maximum number of vertices, lighting and shadows; static batching means that objects that are set to be immovable can be batched when using the same material, but may take up more memory.

\begin{table}[!ht] \centering \small
    \begin{center}
        \caption{DrawCall Optimization}\label{drawcall-test}
        \begin{tabular}{lccc} 
        \hline
            Case & \makecell[c]{Maximum number\\of DrawCalls} &\makecell[c]{Average number\\of DrawCalls} \\
        \hline
            AllDynamic       & 880   & 796.346\\
            SetStatic      & 795 & 663.618\\
        \hline
        \end{tabular}
    \end{center}
\end{table}

We set the objects that do not need to be moved as static objects, and test the applications before and after optimization. AllDynamic refers to the program that did not set static objects before optimization, and SetStatic refers to the program after setting static objects for optimization. The test data is shown in \autoref{drawcall-test}.

We found that after turning on the static object switch of some objects, the maximum number of DrawCalls and the average number of DrawCalls were reduced by 9.7\% and 16.8\% respectively, which greatly optimized the performance of the project.

\section{Experiment on Teaching Effect}

As a virtual reality teaching software, we should also focus on its teaching significance. In order to test the teaching effect of the software, we selected two groups of students to conduct experiments. The experimental contents were the traditional teaching effect test and the VR teaching effect test.

\begin{figure}
    \centering
    \begin{tikzpicture}
        \begin{axis}[
                ybar,
                enlargelimits=.15,
                bar width=.2cm,
                legend style={at={(0.5,1)},
                    anchor=north,legend columns=-1},
                symbolic x coords={Question 1, Question 2, Question 3},
                width=.5\textwidth, height=.4\textwidth,
                ylabel={\%},
                ymin=0,ymax=60,
                xtick=data,
            ]
            \addplot table[x=interval,y=5]{\tradata};
            \addplot table[x=interval,y=4]{\tradata};
            \addplot table[x=interval,y=3]{\tradata};
            \addplot table[x=interval,y=2]{\tradata};
            \addplot table[x=interval,y=1]{\tradata};
            \legend{5, 4, 3, 2, 1}
        \end{axis}
    \end{tikzpicture}
    \caption{Traditional teaching result}\label{traditional-result}
\end{figure}
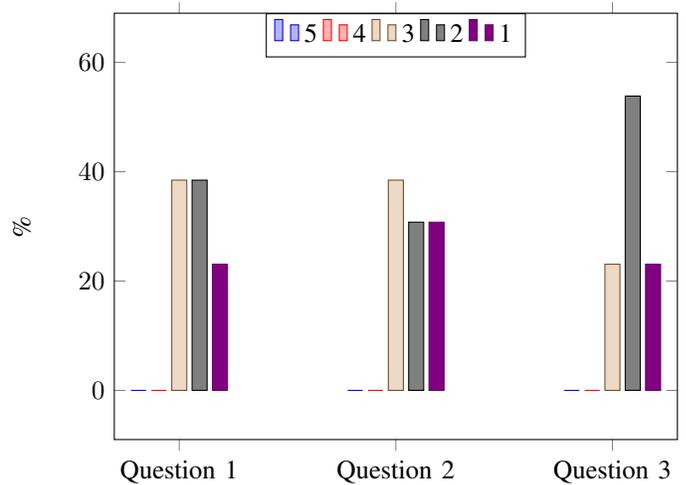

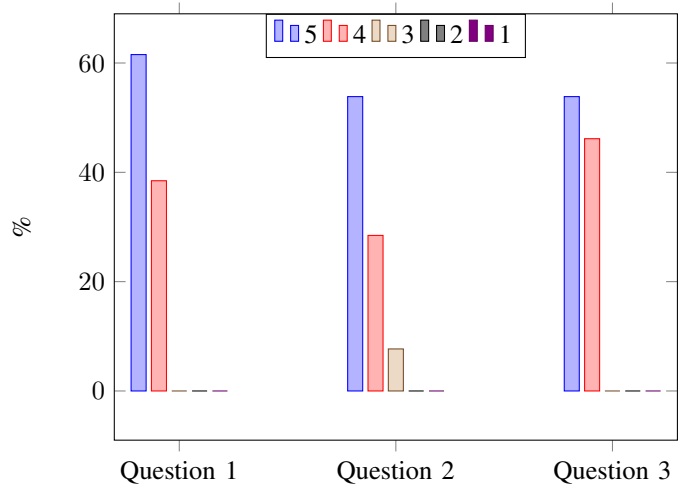
\begin{figure}
    \centering
    \begin{tikzpicture}
        \begin{axis}[
                ybar,
                enlargelimits=.15,
                bar width=.2cm,
                legend style={at={(0.5,1)},
                    anchor=north,legend columns=-1},
                symbolic x coords={Question 1, Question 2, Question 3},
                width=.5\textwidth, height=.4\textwidth,
                ylabel={\%},
                ymin=0,ymax=60,
                xtick=data,
            ]
            \addplot table[x=interval,y=5]{\vrdata};
            \addplot table[x=interval,y=4]{\vrdata};
            \addplot table[x=interval,y=3]{\vrdata};
            \addplot table[x=interval,y=2]{\vrdata};
            \addplot table[x=interval,y=1]{\vrdata};
            \legend{5, 4, 3, 2, 1}
        \end{axis}
    \end{tikzpicture}
    \caption{VR teaching result}
    \label{vr-result}
\end{figure}

The students we selected are all fourth-year students in our school, and they have absolutely no knowledge and experience in engine disassembly and assembly. We selected the first major task of disassembly as the experimental content, and let two groups of students learn by watching the teaching ppt and using this VR teaching software. After learning, let two groups of students evaluate the learning process by filling out questionnaires, and the experimental results obtained are shown in \autoref{traditional-result} and \autoref{vr-result}. The questions are as followed:

\begin{enumerate}
    
    \item \textbf{Question 1:} Do you think you have mastered the disassembly process perfectly?
    \item \textbf{Question 2:} Whether the learning process of this project can increase your interest to continue learning?
    \item \textbf{Question 3:} After completing the study, can you accurately identify the tools you need to use?
\end{enumerate}

We found that students who use VR teaching software to learn are more motivated and more confident in their learning outcomes than students who use traditional methods of theoretical learning. Therefore, our engine disassembly teaching software based on virtual reality has certain educational significance.

\pgfplotstableread[row sep=\\,col sep=&]{
    interval & 5 & 4 & 3 & 2 & 1 \\
    Question 1    & 0  & 0  & 38.46 & 38.46 & 23.08 \\
    Question 2     & 0 & 0  & 38.46 & 30.77 & 30.77 \\
    Question 3   & 0 & 0 & 23.08 & 53.84 & 23.08 \\
    }\tradata

\pgfplotstableread[row sep=\\,col sep=&]{
    interval & 5 & 4 & 3 & 2 & 1 \\
    Question 1    & 61.54  & 38.46  & 0 & 0 & 0 \\
    Question 2     & 53.85 & 28.46  & 7.69 & 0 & 0 \\
    Question 3   & 53.85 & 46.15 & 0 & 0 & 0 \\
    }\vrdata

\section{Summary}

At present, it is difficult for vocational colleges in the automotive industry to carry out practical teaching due to insufficient funding, limited space resources and safety hazards, and teaching is basically based on theoretical teaching. This leads to insufficient motivation for students to learn and difficulty in mastering knowledge to the required degree. Therefore, it makes sense to integrate virtual reality technology into the teaching of the automotive industry. Virtual reality technology has developed rapidly in recent years and has been widely used in various fields of military and medical care, so it is feasible and important to develop an engine disassembly and assembly teaching software based on virtual reality technology.

In this paper, we developed an engine disassembly teaching software from scratch based on the requirements proposed by the company from a teaching scenario using Unity 3D. A clear and distinct system architecture was designed in the software, and the user interface, assembly system, and task system were developed modularly based on the designed system to develop and configure the required functions of the software.

After completing the functional requirements development, we conducted a complete functional and performance test of this software. For functional testing, we designed test cases for the requirement points and conducted multiple rounds of testing to verify the completeness and correctness of the functionality. For performance testing, we used Unity UPR to monitor and record the performance data of the running software, and to tune the unsatisfactory performance data. We reduced cpu time by turning off vertical synchronization to increase frame rate. We reduced the number of DrawCall by turning on static switches and using static batching with the same materials whenever possible, which ultimately resulted in performance data that met the criteria.

Finally, we delivered the product to the company for implementation into practical vocational education applications. We hope that in the future, virtual reality technology can be applied to education more often to achieve the integration of technology and education and cultivate more talents.

\bibliographystyle{IEEEtran}
\bibliography{bibliography}

\begin{thebibliography}{10}
\providecommand{\url}[1]{#1}
\csname url@samestyle\endcsname
\providecommand{\newblock}{\relax}
\providecommand{\bibinfo}[2]{#2}
\providecommand{\BIBentrySTDinterwordspacing}{\spaceskip=0pt\relax}
\providecommand{\BIBentryALTinterwordstretchfactor}{4}
\providecommand{\BIBentryALTinterwordspacing}{\spaceskip=\fontdimen2\font plus
\BIBentryALTinterwordstretchfactor\fontdimen3\font minus
  \fontdimen4\font\relax}
\providecommand{\BIBforeignlanguage}[2]{{%
\expandafter\ifx\csname l@#1\endcsname\relax
\typeout{** WARNING: IEEEtran.bst: No hyphenation pattern has been}%
\typeout{** loaded for the language `#1'. Using the pattern for}%
\typeout{** the default language instead.}%
\else
\language=\csname l@#1\endcsname
\fi
#2}}
\providecommand{\BIBdecl}{\relax}
\BIBdecl

\bibitem{lan2017development}
G.~Lan, J.~Liang, G.~Liu, and Q.~Hao, ``Development of a smart floor for target
  localization with bayesian binary sensing,'' in \emph{2017 IEEE 31st
  International Conference on Advanced Information Networking and Applications
  (AINA)}.\hskip 1em plus 0.5em minus 0.4em\relax IEEE, 2017, pp. 447--453.

\bibitem{xu2019online}
H.~Xu, G.~Lan, S.~Wu, and Q.~Hao, ``Online intelligent calibration of cameras
  and lidars for autonomous driving systems,'' in \emph{2019 IEEE Intelligent
  Transportation Systems Conference (ITSC)}.\hskip 1em plus 0.5em minus
  0.4em\relax IEEE, 2019, pp. 3913--3920.

\bibitem{AiLian2011Vocational}
S.~Ai~Lian, ``The application study of the virtual reality technique in auto
  electronics teaching of higher vocational education(in chinese),''
  \emph{Journal of Human Institute of Science and Technology}, vol.~24, no.~4,
  p.~3, 2011.

\bibitem{2017Application}
W.~Min~Fei, S.~Lin~Yan, Z.~Shuai~Wu, and Z.~Kai, ``Application of virtual
  reality technology in information teaching of automobile repairing
  specialty(in chinese),'' \emph{Journal of Hubei Industrial Polytechnic},
  vol.~30, no.~2, p.~4, 2017.

\bibitem{Tingting2021VR}
L.~Tingting, D.~Nannan, and Q.~Chenyu, ``Research on the application of vr
  technology in vocational education and teaching-based on the flipped
  classroom teaching mode of auto repair major(in chinese),'' \emph{Auto Time},
  no.~20, p.~2, 2021.

\bibitem{Juan2014VR}
L.~Juan, ``Virtual reality and its application in teaching of automobile
  maintenance(in chinese),'' \emph{Xinkecheng Xuexi (shehuizonghe)}, no.~11,
  2014.

\bibitem{lan2018directed}
G.~Lan, M.~Jelisavcic, D.~M. Roijers, E.~Haasdijk, and A.~E. Eiben, ``Directed
  locomotion for modular robots with evolvable morphologies,'' in
  \emph{International Conference on Parallel Problem Solving from
  Nature}.\hskip 1em plus 0.5em minus 0.4em\relax Springer, 2018, pp. 476--487.

\bibitem{DeFu2017Application}
L.~De~Fu, ``Application of 3d virtual simulation technology in automobile
  maintenance teaching(in chinese),'' Ph.D. dissertation, Shenzhen University,
  2017.

\bibitem{lan2022semantic}
G.~Lan, T.~Liu, X.~Wang, X.~Pan, and Z.~Huang, ``A semantic web technology
  index,'' \emph{Scientific Reports}, vol.~12, no.~1, pp. 1--10, 2022.

\bibitem{gao2021neat}
Z.~Gao and G.~Lan, ``A neat-based multiclass classification method with class
  binarization,'' in \emph{Proceedings of the Genetic and Evolutionary
  Computation Conference Companion}, 2021, pp. 277--278.

\bibitem{MinJie2011Research}
L.~Min~Jie, ``Research on virtual simulation training teaching of secondary
  vocational auto repair(in chinese),'' Ph.D. dissertation, Zhejiang University
  of Technology, 2011.

\bibitem{Ferrero1999simulation}
A.~Ferrero and V.~Piuri, ``A simulation tool for virtual laboratory experiments
  in a www environment,'' \emph{IEEE Transactions on Instrumentation and
  Measurement}, vol.~48, no.~3, pp. 741--746, 1999.

\bibitem{su2014case}
C.~Su, W.~Xu, and C.~Feng~Kuang, ``A case study of augmented reality simulation
  system application in a chemistry course,'' \emph{Computers in Human
  Behavior}, vol.~37, p. 31–40, 2014.

\bibitem{Potkonjak2016Virtual}
V.~Potkonjak, M.~Gardner, V.~Callaghan, P.~Mattila, C.~Guetl, V.~M. Petrovic,
  and K.~Jovanovic, ``Virtual laboratories for education in science,
  technology, and engineering: A review,'' \emph{Computers and Education},
  vol.~95, no. Apr., pp. 309--327, 2016.

\bibitem{Siemens2004Connectivism}
G.~Siemens, ``Connectivism: A learning theory for the digital age,''
  \emph{international journal of instructional technology and distance
  learning}, 2004.

\bibitem{Xiao2018Scene}
L.~Xiao, ``Virtual interactive reproduction of historical and cultural scenes
  in northwest china(in chinese),'' Ph.D. dissertation, 2018.

\bibitem{lan2019simulated}
G.~Lan, J.~Chen, and A.~E. Eiben, ``Simulated and real-world evolution of
  predator robots,'' in \emph{2019 IEEE Symposium Series on Computational
  Intelligence (SSCI)}.\hskip 1em plus 0.5em minus 0.4em\relax IEEE, 2019, pp.
  1974--1981.

\bibitem{lan2019evolutionary}
G.~Lan, J.~Chen, and {Eiben, A. E.}, ``Evolutionary predator-prey robot
  systems: From simulation to real world,'' in \emph{Proceedings of the Genetic
  and Evolutionary Computation Conference Companion}, 2019, pp. 123--124.

\bibitem{lan2018real}
G.~Lan, J.~Benito-Picazo, D.~M. Roijers, E.~Dom{\'\i}nguez, and A.~E. Eiben,
  ``Real-time robot vision on low-performance computing hardware,'' in
  \emph{2018 15th international conference on control, automation, robotics and
  vision (ICARCV)}.\hskip 1em plus 0.5em minus 0.4em\relax IEEE, 2018, pp.
  1959--1965.

\bibitem{lan2019evolving}
G.~Lan, L.~De~Vries, and S.~Wang, ``Evolving efficient deep neural networks for
  real-time object recognition,'' in \emph{2019 IEEE Symposium Series on
  Computational Intelligence (SSCI)}.\hskip 1em plus 0.5em minus 0.4em\relax
  IEEE, 2019, pp. 2571--2578.

\bibitem{Chao2017survey}
H.~Chao, T.~Feng, and C.~Ling~Wei, ``Application of vr in education and
  training: A survey(in chinese),'' \emph{Audio Engineering}, vol.~41, no.~11,
  p.~8, 2017.

\bibitem{Feng0AR}
G.~Feng, Z.~De~Yao, L.~Dan, L.~Guan~Fu, Y.~Qian, and Y.~Su~Lian, ``Development
  of ar teaching software for automobile engine construction and maintenance
  based on unity 3d(in chinese).''

\bibitem{KeRong2020Thoughts}
Y.~Ke~Rong and Y.~Li, ``Thoughts on improving the teaching efficiency of engine
  disassembly and assembly training course(in chinese),'' \emph{jiaoyu jiaoxue
  luntan}, vol. No.455, no.~09, pp. 316--317, 2020.

\bibitem{lan2022time}
G.~Lan, J.~M. Tomczak, D.~M. Roijers, and A.~E. Eiben, ``Time efficiency in
  optimization with a bayesian-evolutionary algorithm,'' \emph{Swarm and
  Evolutionary Computation}, vol.~69, p. 100970, 2022.

\bibitem{lan2021learning2}
G.~Lan, M.~De~Carlo, F.~van Diggelen, J.~M. Tomczak, D.~M. Roijers, and A.~E.
  Eiben, ``Learning directed locomotion in modular robots with evolvable
  morphologies,'' \emph{Applied Soft Computing}, vol. 111, p. 107688, 2021.

\bibitem{lan2021learning1}
G.~Lan, M.~van Hooft, M.~De~Carlo, J.~M. Tomczak, and A.~E. Eiben, ``Learning
  locomotion skills in evolvable robots,'' \emph{Neurocomputing}, vol. 452, pp.
  294--306, 2021.

\bibitem{lan2016developmentVR}
G.~Lan, Z.~Luo, and Q.~Hao, ``Development of a virtual reality teleconference
  system using distributed depth sensors,'' in \emph{2016 2nd IEEE
  International Conference on Computer and Communications (ICCC)}.\hskip 1em
  plus 0.5em minus 0.4em\relax IEEE, 2016, pp. 975--978.

\bibitem{lan2016developmentUAV}
G.~Lan, J.~Sun, C.~Li, Z.~Ou, Z.~Luo, J.~Liang, and Q.~Hao, ``Development of
  uav based virtual reality systems,'' in \emph{2016 IEEE International
  Conference on Multisensor Fusion and Integration for Intelligent Systems
  (MFI)}.\hskip 1em plus 0.5em minus 0.4em\relax IEEE, 2016, pp. 481--486.

\end{thebibliography}

\end{document}